\documentclass{elsart}
\usepackage{amssymb}
\usepackage{amsfonts}
\usepackage{epsfig}
\begin{document}
 \noindent\mbox{}  \mbox{} \hfill 
{\sl Published in: \underline{Neurocomputing 63, 125-137 (2004)}}\\
\begin{frontmatter}
\title{Magnification Control in Winner Relaxing Neural Gas}
\author[Kiel]{Jens Christian Claussen} and \author[Leipzig]{Thomas Villmann}
\address[Kiel]{Christian-Albrechts-Universit\"{a}t zu Kiel,
Institut f\"{u}r Theoretische Physik und Astrophysik,
D--24098 Kiel, Leibnizstr.15, Germany}
\address[Leipzig]{Universit\"{a}t Leipzig,
Klinik f\"{u}r Psychotherapie, 
D--04107 Leipzig, Karl--Tauchnitz--Str.25, Germany%
  \\[1ex] \rm 
24 October 2003, final version 2 February, 2004
}
\begin{abstract}
An important goal in neural map learning, 
which can conveniently be accomplished by magnification control,
is to achieve information optimal coding in the
sense of information theory.
In the present
contribution we consider the winner relaxing approach for the neural gas
network. Originally, winner relaxing learning is a slight modification 
of the self-\-or\-ga\-ni\-zing map learning rule that
allows for adjustment of the magnification behavior by an 
\textsl{a priori} chosen control parameter. We transfer this approach to the
neural gas algorithm. The magnification exponent can be
calculated analytically for arbitrary dimension from a continuum theory, and
the entropy of the resulting map is studied numerically conf
irming the theoretical prediction. The influence of a
diagonal term, which can be added without impacting the magnification, 
is studied numerically. This approach to maps of maximal mutual
information is interesting for applications as the winner relaxing term only
adds computational cost of same order and is easy to implement. In
particular, it is not necessary to estimate the generally unknown data
probability density as in other magnification control
approaches.
\end{abstract}
\end{frontmatter}

\thispagestyle{empty}

\section{Introduction}

Neural maps are a widely ranging class of neural vector quantizers which are
commonly used {e.g.} in data visualization, feature extraction, principle
component analysis, image processing, and classification tasks. A well
studied approach is the Neural Gas Network (NG) \cite{martinetz93d}. An
important advantage of the NG is the adaptation dynamics, which minimizes a
potential, in contrast to the self-organizing map (SOM) \cite{Kohonen95a}
frequently used in vector quantization problems.

In the present paper we consider a new control scheme for the \textsl{%
magnification} of the map \cite%
{bishop97a,Claussen2002b,dersch95a,luttrell91a,ritter92a,Villmann2000e}.
Controlling the magnification factor is relevant for many applications in
control theory or robotics, were (neural) vector quantizers are often used
to determine the actual state of the system in a first step, which is an
objective of the control task \cite{Villmann99j,Villmann2000e}. For
instance, in \cite{Villmann2000h} was demonstrated that the application of a
magnification control scheme for the neural gas based classification system
of position and movement state of a robot can reduce the crash probability.
Another area of application is information-theoretically optimal coding of
high-dimensional data as occur in satellite remote sensing image analysis of
hyperspectral images \cite{Merenyi2004a,Villmann2003e} which is, in fact,
the task of equiprobabilistic mapping \cite{van_hulle00a}. Further
applications can be found in medical visualization and classification tasks 
\cite{Wismueller2002a}. Generally, vector quantization according to an
arbitrary $l_{p}$-norm can be related to the problem of magnification
control as it is explained below.

\section{The neural gas network}
The NG maps data vectors $\mathbf{v}$ from a (possibly high-dimensional)
data manifold $\mathcal{D}\subseteq $\textbf{$\mathbb{R}$}$^{d}$ onto a set 
$A$ of neurons $i$, formally written as $\Psi _{\mathcal{D}\rightarrow A}:%
\mathcal{D}\rightarrow A$. Each neuron $i$ is associated with a pointer $%
\mathbf{w}_{i}\in $\textbf{$\mathbb{R}$}$^{d}$ also called weight vector,
or codebook vector.
All weight vectors establish the set $\mathbf{W}=\left\{ \mathbf{w}%
_{i}\right\} _{i\in A}$. The mapping description is a winner take all rule,
i.e. a stimulus vector $\mathbf{v}\in \mathcal{D}$ is mapped onto the neuron 
$s\in A$ the pointer $\mathbf{w}_{s}$ of which is closest to the actually
presented stimulus vector $\mathbf{v}$, 
\begin{equation}
\Psi _{\mathcal{D}\rightarrow A}:\mathbf{v}\mapsto s\left( \mathbf{v}\right)
=\mathop{\rm argmin}_{i\in A}\left\Vert \mathbf{v}-\mathbf{w}_{i}\right\Vert.  
\label{argmin}
\end{equation}%
The neuron $s\left( \mathbf{v}\right) $ is called \textsl{winner neuron}.
The set 
\begin{equation}
\Omega _{i}=\left\{ \mathbf{v\in }\mathcal{D}|\Psi _{\mathcal{D}\rightarrow
A}\left( \mathbf{v}\right) =i\right\}
\end{equation}%
is called (masked) receptive field of the neuron $i$.

During the adaptation process a sequence of data points $\mathbf{v}\in 
\mathcal{D}$ is presented to the map with respect to the stimuli
distribution $P\left( \mathcal{D}\right) $. Each time the currently most
proximate neuron $s$ according to (\ref{argmin}) is determined, and the
pointer $\mathbf{w}_{s}$ as well as all pointers $\mathbf{w}_{i}$ of neurons
in the neighborhood of $\mathbf{w}_{s}$ are shifted towards $\mathbf{v}$,
according to 
\begin{equation}
\bigtriangleup \mathbf{w}_{i}=\epsilon h_{\lambda }\left( i,\mathbf{v},%
\mathbf{W}\right) \left( \mathbf{v}-\mathbf{w}_{i}\right) .
\label{allg_lernen}
\end{equation}%
The property of \textquotedblleft being in the neighborhood of $\mathbf{w}%
_{s}$\textquotedblright\ is represented by a neighborhood function $%
h_{\lambda }\left( i,\mathbf{v},\mathbf{W}\right) $. The neighborhood
function is defined as 
\begin{equation}
h_{\lambda }\left( i,\mathbf{v},\mathbf{W}\right) =\exp \left( -\frac{%
k_{i}\left( \mathbf{v},\mathbf{W}\right) }{\lambda }\right) ,  \label{h_trn}
\end{equation}%
where $k_{i}\left( \mathbf{v},\mathbf{W}\right)$ 
is defined as 
the number of
pointers $\mathbf{w}_{j}$ for which the relation 
$\left\Vert \mathbf{v}-\mathbf{w}_{j}\right\Vert
 \leq \left\Vert \mathbf{v}-\mathbf{w}_{i}\right\Vert$ 
is valid, i.e. $k_{i}\left( \mathbf{v},\mathbf{W}\right) ~$%
\ is the winning rank \cite{martinetz93d}. In particular, for the winning
neuron $s$ we have $h_{\lambda }\left( s,\mathbf{v},\mathbf{W}\right) =1.0$.
We remark that in contrast to the SOM the neighborhood function is evaluated
in the input space. Moreover, the adaptation rule for the weight vectors in
average follows a potential dynamics \cite{martinetz93d}.

The \textsl{magnification} of the trained map reflects the relation between
the data density $P\left( \mathcal{D}\right) $ and the density $\rho $ of
the weight vectors \cite{ritter91c}. For the NG the relation 
\begin{equation}
\rho \left( \mathbf{w}\right) \propto P\left( \mathcal{D}\right) ^{\alpha _{%
\mathsf{NG}}}  \label{density_power_law}
\end{equation}
with 
\begin{equation}
\alpha _{\mathsf{NG}}=\frac{d}{d+2}
\end{equation}
has been derived \cite{martinetz93d}. The exponent $\alpha _{\mathsf{NG}}$
is called \textsl{magnification factor}. For the NG it depends on the \emph{%
intrinsic} dimensionality $d$ of the data which can be numerically
determined by several methods \cite%
{bruske95a,Camastra2003a,Camastra2001a,Grassberger83a,Takens85a}. For
simplicity we further require that the 
(embedding) data dimension is the intrinsic one.

Generally, the information transfer is not independent of the magnification
of the map \cite{Zador82a}. It is known that for a vector quantizer (or a
neural map in our context) with optimal information transfer the relation $%
\alpha =1$ holds. Otherwise, a vector quantizer which minimizes the mean
distortion error 
\begin{equation}
E_{\gamma }=\int_{\mathcal{D}}\left\Vert \mathbf{w}_{s}-\mathbf{v}%
\right\Vert ^{\gamma }P\left( \mathbf{v}\right) d\mathbf{v}
\label{vq_energy}
\end{equation}%
has the magnification factor 
\begin{equation}
\alpha =\frac{{d}}{{d+\gamma }}  \label{vq_mgn}
\end{equation}%
with $\mathbf{v\in }\mathcal{D}\subseteq $ \textbf{$\mathbb{R}$}$^{d}$, i.e.
the magnification of a vector quantizer is directly related to the
minimization of the description error according to a certain $l_{p}$-norm 
\cite{Zador82a}. Hence, the NG minimizes the usual $E_{2}$ distortion error.

We now address the question how to extend the NG to achieve an \textsl{a
priori} chosen optimization goal, {i.e.} an \textsl{a priori} chosen
magnification factor. 

\section{Controlling the magnification in NG\label{sectioncontroltrn}}
For the SOM several methods exist to control the magnification
of the map. The first approach to influence the magnification of a learning
vector quantizer, proposed in \cite{desieno88} is called the \emph{mechanism
of conscience}. For this purpose a bias term is added in the winner rule 
(\ref{argmin}):
\begin{equation}
\Psi _{\mathcal{D}\rightarrow A}:\mathbf{v}\mapsto s\left( 
\mathbf{v}\right)=
\mathop{\rm argmin}_{i\in A}
\left( \left\Vert \mathbf{v}-\mathbf{w}%
_{i}\right\Vert -\gamma \left( \frac{1}{N}-p_{i}\right) \right)
\label{argmin_desieno}
\end{equation}%
where $p_{i}$ is the actual winning probability of the neuron $i$ and $%
\gamma $ is a balance factor. Hence, the winner determination is influenced
by this modification. The algorithm should converge such that the winning
probabilities of all neurons are equalized. This is related to a
maximization of the entropy and consequently the resulting magnification is
equal to unity. 
However, as pointed out by \cite{van_hulle00a}, adding a
conscience algorithm to the SOM does not equate to equiprobabilistic
mapping, in general. Only for {\itshape very high
dimensions}, a minimum distortion quantizer (such as the conscience
algorithm) approaches an equiprobable quantizer (\cite{van_hulle00a} - page
93). Further, an arbitrary magnification cannot be achieved by this
mechanism. Moreover, numerical studies of the algorithm have shown
instabilities \cite{van_hulle00a}. 
To control the magnification, a local learning parameter 
was introduced \cite{bauer96a} into the usual SOM-learning scheme.
The now localized learning allows in principle an arbitrary magnification. Other
authors proposed variants which lead more away from the original SOM by
kernel methods \cite{van_hulle00a} or statistical approaches \cite{Linsker89a}.

For the NG a solution of the magnification control problem can be realized
by introducing an adaptive \emph{local learning} step size $\epsilon
_{s\left( \mathbf{v}\right) }$ \cite{Villmann2000e} according to the above
mentioned approach for SOM \cite{bauer96a}. Then, the new \emph{localized}
learning rule reads as 
\begin{equation}
\bigtriangleup \mathbf{w}_{i}=\epsilon _{s\left( \mathbf{v}\right)
}h_{\lambda }\left( i,\mathbf{v},\mathbf{W}\right) \left( \mathbf{v}-\mathbf{%
w}_{i}\right)  \label{trn_local_lernen}
\end{equation}%
with the local learning parameters $\epsilon _{i}=\epsilon \left( \mathbf{w}%
_{i}\right) $ depending on the stimulus density $P$ at the position of the
weight vectors $\mathbf{w}_{i}$ via 
\begin{equation}
\left\langle \epsilon _{i}\right\rangle =\epsilon _{0}P\left( \mathbf{w}%
_{i}\right) ^{m}.
\end{equation}%
The brackets $\left\langle \ldots \right\rangle $ denote the average in
time, and $s\left( \mathbf{v}\right) $ is the best--matching neuron with
respect to (\ref{argmin}). Note, that the local learning rate $\epsilon
_{s\left( \mathbf{v}\right) }$ of the winning neuron is applied in the
adaptation step (\ref{trn_local_lernen}) for each neuron. This approach
finally leads to the new magnification law 
\begin{equation}
\alpha ^{\prime }=\alpha _{NG}\cdot \left( m+1\right)
\end{equation}%
which is a modification of the old one. Hence, the parameter $m$ plays the
role of a control parameter.

However, in real applications one has to estimate the generally unknown data
distribution $P$. Usually this is done by estimation of the volume of the
receptive fields and the firing rates \cite{bauer96a,Villmann98e}. This may
lead to numerical instabilities of the control mechanism \cite%
{Villmann97n,van_hulle00a,Villmann99j}. Therefore, an alternative control
mechanism is demanded.

Recently, a new approach for magnification control of the SOM was introduced 
\cite{Claussen2002a,Claussen2003a} which avoids the $P$-estimation problem.
The respective approach is a generalization of a 
modification of the usual SOM \cite{kohonen91c}. It is called (generalized)
Winner Relaxing SOM (WRSOM) \cite{Claussen2002a,Claussen2003a}. 
In winner relaxing SOM an
additional term occurs in weight vector update for the winning neuron,
implementing a relaxing behavior. The relaxing force is a weighted sum of
the difference between the weight vectors and the input according to their
distance rank. The relaxing term was originally introduced 
in \cite{kohonen91c} to obtain a learning dynamic
for SOM according to an average reconstruction error including the effect of
shifting Voronoi borders.

It was shown that the generalized winner relaxing mechanism applied in WRSOM
can be used for magnification control in SOM, too \cite{Claussen2002a}.
Thereby, the winner relaxing approach provides a magnification control
scheme for SOM which is \textsl{independent} of the shape of the data
distribution only depending on parameters of the winner relaxing term.

\subsection{The winner relaxing neural gas}
We now transfer the generalized winner relaxing approach for SOM to the NG
and consider its influence on the magnification. In complete analogy to the
WRSOM we add a general winner relaxing term $R\left( \xi ,\kappa \right) $
to the usual NG-learning dynamic (\ref{allg_lernen}). Then the weight update
reads as 
\begin{equation}
\bigtriangleup \mathbf{w}_{i}=\epsilon h_{\lambda }\left( i,\mathbf{v},%
\mathbf{W}\right) \left( \mathbf{v}-\mathbf{w}_{i}\right) +R\left( \xi
,\kappa \right),  \label{winner_relaxing_learning}
\end{equation}%
whereby the winner relaxing term is defined as 
\begin{equation}
R\left( \xi ,\kappa \right) =\left( \xi +\kappa \right) \left( \mathbf{v}-%
\mathbf{w}_{i}\right) \delta _{is}-\kappa \delta _{is}\sum_{j}h_{\lambda
}\left( j,\mathbf{v},\mathbf{W}\right) \left( \mathbf{v}-\mathbf{w}%
_{j}\right)
\end{equation}%
depending on the additional weighting parameters $\xi$ and $\kappa$. We
refer to this algorithm as the \textsl{winner relaxing NG} (WRNG). The
original winner relaxing term described in \cite{kohonen91c} is obtained for
the special parameter choice $\xi =0,\kappa =\frac{1}{2}$. Note, that the
relaxing term only contributes to the winner weight vector update as in the
original approach.

\subsection{Derivation of the Magnification for WRNG}
We now derive a relation between the densities $\rho $ and $P$ in analogy to 
\cite{martinetz93d} for the winner relaxing learning (\ref%
{winner_relaxing_learning}). The procedure is very similar as in \cite%
{martinetz93d,Villmann2000e}. The average change $\left\langle
\bigtriangleup \mathbf{w}_{i}\right\rangle $ for the winner relaxing NG
learning rule (\ref{winner_relaxing_learning}) is
\begin{eqnarray}\nonumber
\left\langle \bigtriangleup \mathbf{w}_{i}\right\rangle &=&\int P\left( 
\mathbf{v}\right) h_{\lambda }\left( i,\mathbf{v},\mathbf{W}\right) \left( 
\mathbf{v-w}_{i}\right) +\left( \xi +\kappa \right) {\cdot }\delta
_{is}\cdot \left( \mathbf{v-w}_{i}\right) 
\\&&-\delta _{is}\kappa
\sum_{j}h_{\lambda }\left( j,\mathbf{v},\mathbf{W}\right) \left( \mathbf{v-w}%
_{j}\right) d\mathbf{v.}  \label{average_change}
\end{eqnarray}
We now consider the equilibrium state, i.e. $\left\langle \bigtriangleup 
\mathbf{w}_{i}\right\rangle =0$.

For this purpose, we first separate the integral (\ref{average_change}) into 
\begin{equation}
\left\langle \bigtriangleup \mathbf{w}_{i}\right\rangle =I_{1}+I_{2}+I_{3}
\end{equation}%
with 
\begin{equation}
I_{1}=\int P\left( \mathbf{v}\right) h_{\lambda }\left( i,\mathbf{v},\mathbf{%
W}\right) \left( \mathbf{v-w}_{i}\right) d\mathbf{v,}  \label{I1_integral}
\end{equation}%
\begin{equation}
I_{2}=\int P\left( \mathbf{v}\right) \left( \xi +\kappa \right) {\cdot }%
\delta _{is}\cdot \left( \mathbf{v-w}_{i}\right) d\mathbf{v}
\label{I2_integral}
\end{equation}%
and 
\begin{equation}
I_{3}=-\int P\left( \mathbf{v}\right) \delta _{is}\kappa \sum_{j}h_{\lambda
}\left( j,\mathbf{v},\mathbf{W}\right) \left( \mathbf{v-w}_{j}\right) d%
\mathbf{v}  \label{I3_integral}
\end{equation}%
The integral $I_{1}$ is the usual one according to the NG dynamics whereas $%
I_{2}$, $I_{3}$ are related to the winner relaxing scheme. In the following
we treat each integral in a separate manner. Thereby we always assume a
continuum approach, i.e. the index $i$ becomes continuous. Hence, for a
given input $\mathbf{v}$ one can find an optimal $\mathbf{w}_{s}$ 
fulfilling even 
$\mathbf{w}_{s}=\mathbf{v}$ \cite{ritter92a}.

Doing so, the $I_{2}$-integral vanishes in the (first order) continuum limit
because the integration over $\delta _{is}$ only contributes for $\mathbf{w}%
_{s}$, but in this case $\left( \mathbf{v-w}_{s}\right) =0$ holds.

We now pay attention to the $I_{3}$-integral: The continuum assumption made
above allows a turn over from sum $\sum_{j}h_{\lambda }\left( \mathbf{w}_{j},%
\mathbf{v},\mathbf{W}\right) \left( \mathbf{v-w}_{j}\right) $ to the
integral form $\int h_{\lambda }\left( \mathbf{w},\mathbf{v},\mathbf{W}%
\right) \left( \mathbf{v-w}\right) d\mathbf{w}$ in (\ref{I3_integral}). The
further treatment is in complete analogy to the derivation of the
magnification in the usual NG \cite{martinetz93d}. Let $\mathbf{r}$ be the
difference vector%
\begin{equation}
\mathbf{r}=\mathbf{v}-\mathbf{w}_{i}
\end{equation}%
The winning rank $k_{i}\left( \mathbf{v},\mathbf{W}\right) $ only depends on 
$\mathbf{r}$, therefore we introduce the new variable 
\begin{equation}
\mathbf{x}\left( \mathbf{r}\right) =\mathbf{\hat{r}\cdot k}_{i}\left( 
\mathbf{r}\right) ^{\frac{1}{d}}
\end{equation}%
which can be assumed as monotonously increasing with $\left\Vert \mathbf{r}%
\right\Vert $. Thus, the inverse $\mathbf{r}\left(\mathbf{x}\right)$ exists
and we can rewrite the $I_{3}$-integral (\ref{I3_integral}) into 
\begin{equation}
I_{3}=\int P\left( \mathbf{v}\right) \delta _{is}\kappa \left[ \int
h_{\lambda }\left( \mathbf{x}\right) \cdot \mathbf{r}\left( \mathbf{x}%
\right) \cdot \mathbf{J}\left( \mathbf{x}\right) d\mathbf{x}\right] d\mathbf{%
v}
\end{equation}%
with the $d\times d$--Jacobian--matrix 
\begin{equation}
\mathbf{J}\left( \mathbf{x}\right) =\det \left( \frac{{\partial r_{k}}}{{%
\partial x_{l}}}\right) .
\end{equation}%
$I_{3}$ only contributes to $\left\langle \bigtriangleup \mathbf{w}%
_{i}\right\rangle $ for the winning weight (realized by $\delta _{is}$),
i.e., for $\mathbf{w}_{i}=\mathbf{w}_{s}$ which is equal to $\mathbf{v}$
according to the continuum approach. Hence, the integration over $\mathbf{v}$
yields%
\begin{equation}
I_{3}=\kappa P\left( \mathbf{w}_{i}\right) \cdot \int h_{\lambda }\left( 
\mathbf{x}\right) \cdot \mathbf{r}\left( \mathbf{x}\right) \cdot \mathbf{J}%
\left( \mathbf{x}\right) d\mathbf{x}  \label{I_3_equation}
\end{equation}%
If $h_{\lambda }\left( \mathbf{k}_{i}\left( \mathbf{r}\right) \right) $
rapidly decreases to zero with increasing $\mathbf{r}$, we can replace the
quantities $\mathbf{r}\left( \mathbf{x}\right) $, $\mathbf{J}\left( \mathbf{x%
}\right) $ by the first terms of their respective Taylor expansions around
the point $\mathbf{x}=0$ neglecting higher derivatives. We obtain 
\begin{equation}
\mathbf{x}\left( \mathbf{r}\right) =\mathbf{r}\left( \tau _{d}\rho \left( 
\mathbf{w}_{i}\right) \right) ^{\frac{1}{d}}\left( 1+\frac{\mathbf{r}\cdot
\partial _{\mathbf{r}}\rho \left( \mathbf{w}_{i}\right) }{d\cdot \rho \left( 
\mathbf{w}_{i}\right) }+\mathcal{O}\left( \mathbf{r}^{2}\right) \right)
\end{equation}%
which corresponds to 
\begin{equation}
\mathbf{r}\left( \mathbf{x}\right) =\mathbf{x}\left( \tau _{d}\rho \left( 
\mathbf{w}_{i}\right) \right) ^{-\frac{1}{d}}\left( 1-\left( \tau _{d}\rho
\left( \mathbf{w}_{i}\right) \right) ^{-\frac{1}{d}}\cdot \frac{\mathbf{x}%
\cdot \partial _{\mathbf{r}}\rho \left( \mathbf{w}_{i}\right) }{d\cdot \rho
\left( \mathbf{w}_{i}\right) }+\mathcal{O}\left( \mathbf{x}^{2}\right)
\right)
\end{equation}%
with $\tau _{d}={\pi ^{\frac{d}{2}}}/{\Gamma \left( \frac{d}{2}+1\right) }$
as the volume of a $d$--dimensional unit sphere \cite{martinetz93d}.
Further, 
\begin{eqnarray}
\mathbf{J}\left( \mathbf{x}\right) &=&\left( \mathbf{J}\left( 0\right) +x_{k}%
\frac{\partial \mathbf{J}}{\partial x_{k}}+\ldots \right) \\
&=&\left( \tau _{d}\cdot \rho \right) ^{-1}\left( 1-\left( \tau _{d}\cdot
\rho \right) ^{-\frac{1}{d}}\left( 1+\frac{1}{d}\right) \cdot \mathbf{x}%
\cdot \frac{\partial _{\mathbf{r}}\rho }{\rho }\right) +\mathcal{O}\left(
x^{2}\right)
\end{eqnarray}%
and, hence,%
\begin{equation}
\left. \frac{\partial \mathbf{J}}{\partial \mathbf{x}}\right\vert _{\mathbf{x%
}=0}=-\left( \tau _{d}\cdot \rho \right) ^{-\left( 1+\frac{1}{d}\right) }%
\frac{\partial _{\mathbf{r}}\rho }{\rho }.
\end{equation}%
Therefore, the integral in equation (\ref{I_3_equation}) can be rewritten as%
\begin{eqnarray}
I_{3} &=&\epsilon ^{\prime }\kappa P\left( \tau _{d}\cdot \rho \right) ^{-%
\frac{1}{d}}{\displaystyle\int_{\mathcal{D}}}h_{\lambda }\left( \mathbf{x}%
\right) \cdot \mathbf{x}\cdot  \label{Taylor_integral} \\
&&\cdot \left( \left( \tau _{d}\cdot \rho \right) ^{-1}-\left( 1+\frac{1}{d}%
\right) \left( \tau _{d}\cdot \rho \right) ^{-\left( 1+\frac{1}{d}\right)
}\cdot \mathbf{x}\cdot \frac{\partial _{\mathbf{r}}\rho }{\rho }+\ldots
\right)  \nonumber \\
&&\cdot \left( 1-\left( \tau _{d}\cdot \rho \right) ^{-\frac{1}{d}}\cdot 
\mathbf{x}\cdot \frac{\partial _{\mathbf{r}}\rho }{d\cdot \rho }+\ldots
\right) \,\,d\mathbf{x}  \nonumber
\end{eqnarray}

$\allowbreak $ The integral terms in (\ref{Taylor_integral}) of odd order in 
$\mathbf{x}$ vanish because of the rotational symmetry of $h_{\lambda
}\left( \mathbf{x}\right) $. Then (\ref{I_3_equation}) yields, neglecting
terms in higher order in $\mathbf{x}$,

\begin{equation}
I_{3}=\epsilon ^{\prime }\kappa P\frac{d+2}{d}\frac{\partial _{\mathbf{r}%
}\rho }{\rho }  \label{I_3_solution}
\end{equation}%
with%
\begin{equation}
\epsilon ^{\prime }=\frac{\epsilon _{0}}{\left( \tau _{d}\cdot \rho \right)
^{\frac{2+d}{d}}}{\displaystyle\int_{\mathcal{D}}}h_{\lambda }\left( \mathbf{%
x}\right) \cdot \left\Vert \mathbf{x}\right\Vert ^{2}\,\,d\mathbf{x.}
\end{equation}

It remains to consider the $I_{1}$-integral. As mentioned above, it is
identical to the averaged adaptation of the usual NG. Hence, the treatment
can be taken from there and we get 
\begin{equation}
I_{1}=\epsilon ^{\prime }\left( \partial _{\mathbf{r}}P-P\cdot \frac{d+2}{d}%
\cdot \frac{\partial _{\mathbf{r}}\rho }{\rho }\right)  \label{I_1_solution}
\end{equation}%
as an equivalent equation \cite{martinetz93d}.

Taking together (\ref{I_1_solution}) and (\ref{I_3_solution}), the
stationary solution of (\ref{winner_relaxing_learning}) is given by 
\begin{equation}
\left\langle \bigtriangleup \mathbf{w}_{i}\right\rangle =0=\partial _{%
\mathbf{r}}P-P\cdot \frac{d+2}{d}\cdot \frac{\partial _{\mathbf{r}}\rho }{%
\rho }+P\kappa \frac{d+2}{d}\frac{\partial _{\mathbf{r}}\rho }{\rho }
\end{equation}%
This differential equation roughly has the same form as the one for the
usual Neural Gas (\ref{I_1_solution}). Its solution is given by 
\begin{equation}
\rho \varpropto P^{{\alpha }_{\mathsf{WRNG}}}
\end{equation}%
with the exponent 
\begin{equation}
{\alpha }_{\mathsf{WRNG}}=\frac{1}{1-\kappa }\frac{d}{d+2}
\end{equation}%
being the magnification factor. Hence, the magnification factor of the WRNG
can be described also in terms of the magnification of the usual neural gas 
\begin{equation}
{\alpha }_{\mathsf{WRNG}}=\frac{1}{1-\kappa }\alpha _{\mathsf{NG}}
\end{equation}%
Note, that the parameter $\xi $ of the winner relaxing term $R\left( \xi
,\kappa \right) $ does not influence the magnification.

\subsection{Discussion of the theoretical result and comparison with winner
relaxing SOM}

\noindent Two direct observations can be immediately made: Firstly, the
magnification exponent appears to be independent of the additional diagonal
term (controlled by $\xi $) for the winner which is in agreement with the
WRSOM result \cite{Claussen2002a}. Therefore $\xi =0$ again is the usual
setting in WRNG for magnification control. Secondly, by adjusting $\kappa $
appropriately, the magnification exponent can be adjusted, {e.g.} to the
most interesting case of maximum mutual information \cite%
{linsker87a,Zador82a}. Maximum mutual information, which corresponds to
optimal information transfer, is obtained when magnification equals the unit 
\cite{brause92a,brause94a}. Hence, we have for this case the optimum value 
\begin{equation}
\kappa _{\mathsf{opt}}=\frac{2}{d+2}. 
\end{equation}
If the same stability borders $|\kappa |=1$ of the WRSOM also are valid
here, one can expect to increase the NG exponent by positive values of $%
\kappa $, or to lower the NG exponent by a factor $1/2$ for $\kappa =-1$. In
contrast to the Winner Enhancing SOM, where the relaxing term has to be
inverted ($\kappa <0$) to increase the magnification exponent, for the
neural gas positive values of $\kappa $ are required to increase the
magnification exponent. However, the magnification factor still remains
dependent on the generally unknown (intrinsic) dimension of the data. If
this dimension is known, the parameter $\kappa $ can be set \textsl{a priori}
to obtain a neural gas of maximal mutual information. In this approach it is
not necessary to keep track of the local reconstruction errors and firing
rate for each neuron to adjust a local learning rate. Possibilities for
estimating the intrinsic dimension are the well-known
Grassberger-Procaccia-analysis \cite{Grassberger83a} or the neural network
approach using again a NG \cite{bruske95a}.

However, one has to be cautious when transferring the $\lambda \rightarrow
{}0$ result obtained above (which would require to increase the number of
neurons as well) to a realistic situation where a decrease of $\lambda $
with time will be limited to a final finite value to avoid the stability
problems found in \cite{Villmann97n}. If the neighborhood length in SOM is
kept small but fixed for the limit of fine discretization, the neighborhood
function of the second but one winner will again be of order 1 (as for the
winner). For the NG however the neighborhood is defined by the rank list. As
the winner is not present in the $I_{2}+I_{3}$ integral, all terms share the
factor $\mathrm{e}^{-\lambda }$ by $h_{\lambda }(k)=\mathrm{e}^{-\lambda
}h_{\lambda }(k-1)$ which indicates that in the discretized algorithm $%
\kappa $ has to be rescaled by $\mathrm{e}^{+\lambda }$ to agree with the
continuum theory.\footnote{In particular, 
for a finite $\lambda $ the maximum coefficient $h_{\lambda}$
that contributes to the $I_{2}+I_{3}$ integral is given by the prefactor of
the second but one winner, which is given by $\mathrm{e}^{\lambda }$.}

\section{Numerical results}
\begin{figure}[phtb]
\epsfig{file=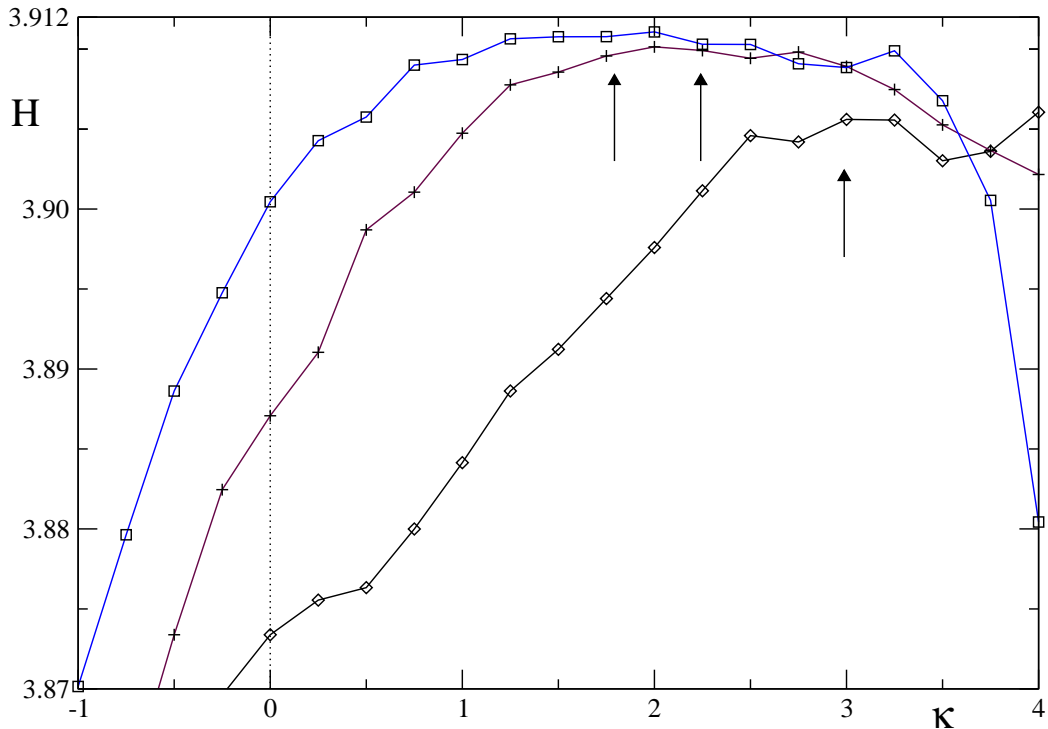,width=\columnwidth}
\caption{Plot of the entropy $H$
according (\protect\ref{entropy}) curves for varying values of $\protect%
\kappa $ for one- $(\diamond )$, two- \mbox{\small\footnotesize
$(+)$}, and
three-dimensional \mbox{\small\footnotesize
$(\Box)$} data. The entropy has
the maximum $\ln (50)\simeq 3.912$ if the magnification equals unity 
\protect\cite{Zador82a}. The arrows indicate the rescaled $\protect\kappa _{%
\mathsf{opt}}$-values for the respective data dimensions.
\label{fig_entropy_results}}
\end{figure}

A numerical study shows how the winner-relaxing mechanism is
able to control the magnification for optimization of the mutual information
of a map generated by the WRNG. Using a standard setup as in \cite%
{Villmann97n} of $N=50$ Neurons and $10^{7}$ training steps with a
probability density $P(x_{1}\ldots x_{d})=\prod_{i}\sin (\pi x_{i})$, with
fixed $\lambda =1.5$ and $\epsilon $ decaying from $0.5$ to $0.05$, the
entropy of the resulting map computed for an input dimension of $1$, $2$ and 
$3$ is plotted in Fig.~\ref{fig_entropy_results}. Thereby, the entropy is
computed using the winning probabilty $p_{i}$ of the neurons:%
\begin{equation}
H=-\sum_{i=1}^{N}p_{i}\ln \left( p_{i}\right)   \label{entropy}
\end{equation}%

The entropy shows a
dimension-dependent maximum approximately at $\kappa =\frac{2}{d+2}\mathrm{e}%
^{\lambda }$. The scaling of the position of the entropy maximum with input
dimension is in agreement with the continuum theory, as well as the
prediction of the opposite sign of $\kappa $ that has to be taken to
increase mutual information. Our numerical investigation indicates that the
above discussed prefactor, in fact, has to be taken in account for finite $%
\lambda $ and a finite number of neurons. We obtain, within a broad range
around the optimal $\kappa $ the entropy is close to the maximum $%
\sum_{i=1}^{N}P_{i}\ln (P_{i})=\ln (N)$ given by information theory.

\begin{figure}[phtb]
\epsfig{file=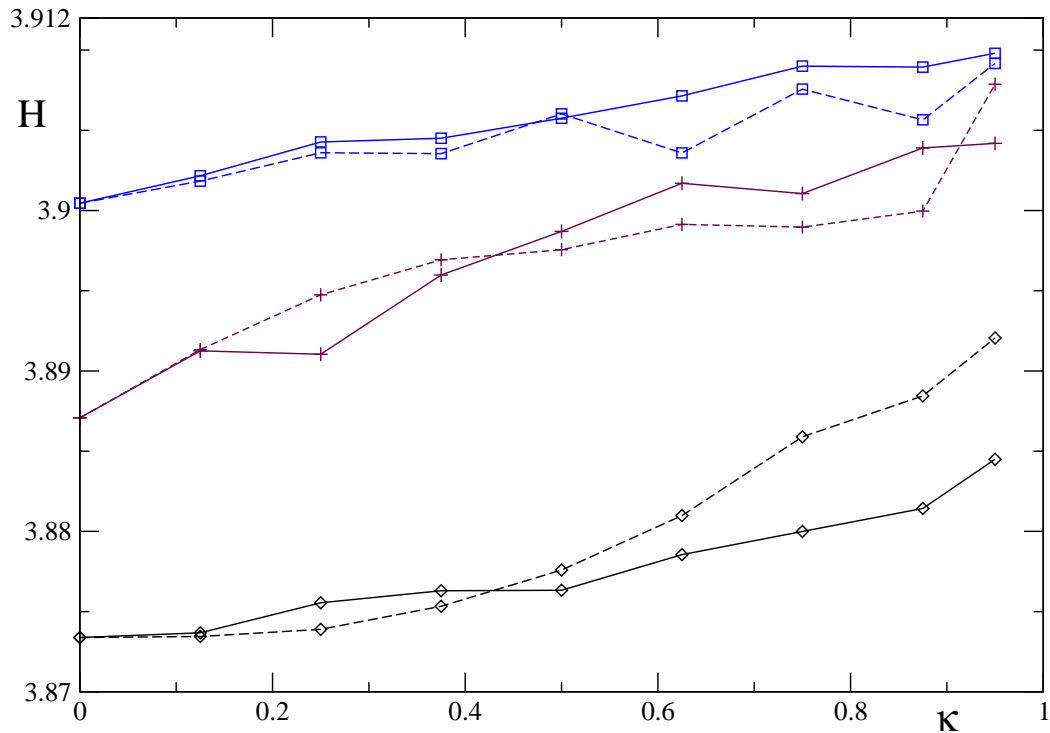,width=\columnwidth}
\caption{Comparison of the entropies curves for different $%
\protect\kappa $-values for $\protect\xi =0$ (straight) and $(\protect\xi +%
\protect\kappa )=0$ (dashed) with respect to one- $(\diamond )$, two- 
\mbox{\small\footnotesize
$(+)$}, and three-dimensional \mbox{\small%
\footnotesize
$(\Box)$} data.
\label{fig_independence_result}}
\end{figure}

In a second numerical study we investigate the influence of the additional
diagonal term (controlled by $\xi $) for the winner. Already for the WRSOM
the magnif\hspace*{0.03em}{}ication exponent is independent of this diagonal
term \cite{Claussen2002a}. In the respective derviation ($I_{2}$-integral (%
\ref{I2_integral})) only first order approximations were used. Otherwise, $%
I_{2}$ may contribute in higher orders. To verify that the contribution of
an additionally added diagonal term is marginal, the entropy was calculated
both for $\xi =0$ and $(\kappa +\xi )=0$ \cite{Villmann2003i}. However, no
influence on the entropy was found for the choice $\kappa +\xi =0$ instead
of $\xi =0$. (Fig~\ref{fig_independence_result}). More pronounced is the inf%
\hspace*{0.03em}{}luence of the diagonal term on stability; according to the
larger pre\-fac\-tor no stable behavior has been found for $|\xi |\geq {}1$,
therefore $\xi =0$ is the recommended setting.

\section{Conclusions}
We introduced a winner-relaxing term in neural gas algorithm to obtain a
winner-relaxing neural gas with the possibility of magnification control.
The winner relaxing scheme is adopted from winner-relaxing SOM. The new
controlling scheme offers a method which is independent on the explicit
knowledge of the generally unknown data distribution which is an advantage
in comparison to the earlier presented neural gas with localized learning
for magnification control. In particular, we avoid the difficult
determination of the data probability density by estimation of the volume of
the receptive fields of the neuron and the firing rate. Numerical
simulations show the abilities of the proposed algorithm.

\textbf{Acknowledgements:} The authors want to thank Th.$\!$ Martinetz 
and H.$\!$ Ritter for detailed comments and intensive discussions.


\end{document}